\documentclass[aip,jcp,author-year,preprint,onecolumn,groupedaddress,amsmath,amssymb,floatfix]{revtex4-1}
\usepackage{graphicx}

\begin{document}


\title{Physical aging and relaxation of residual stresses in a colloidal glass following flow cessation}


\author{Ajay Singh Negi}
\email[]{ajay.negi@yale.edu}
\affiliation{Department of Chemical Engineering, Yale University,
New Haven CT 06511}
\author{Chinedum O. Osuji}
\email[]{chinedum.osuji@yale.edu}
\affiliation{Department of Chemical Engineering, Yale University,
New Haven CT 06511}


\date{\today}


\begin{abstract}

Dilute Laponite suspensions in water at low salt concentration form repulsive colloidal glasses which display physical aging. This
phenomenon is still not completely understood and in particular,
little is known about the connection between the flow history, as a
determinant of the initial state of the system, and the subsequent
aging dynamics. Using a stress controlled rheometer, we perform
stress jump experiments to observe the elastic component of the flow
stress that remains on cessation of flow or flow quenching. We
investigate the connection between the dynamics of these residual
stresses and the rate of physical aging upon quenching from
different points on the steady state flow curve. Quenching from high
rates produces a fluid state, $G''>G'$, with small, fast relaxing
residual stresses and rapid, sigmoidal aging of the complex modulus.
Conversely, quenching from lower shear rates produces increasingly
jammed states featuring slowly relaxing stresses and a slow increase
of the complex modulus with system age. Flow cessation from a fixed
shear rate with varying quench durations shows that slower quenches
produce smaller residual stresses at short times which relax at long
times by smaller extents, by comparison with faster quenches. These
smaller stresses are correlated with a higher modulus but slower
physical aging of the system. The characteristic time for the
residual stress relaxation scales inversely with the quench rate.
This implies a frustrated approach to any ideal stress-free state
that succinctly reflects the frustrated nature of these glassy
systems.

\end{abstract}

\maketitle

\section{\label{introduction}Introduction}

Many complex fluids show some degree of time dependence in their
properties. Such displays of thixotropy are well known in colloidal
dispersions for example, both under attractive and repulsive
inter-particle potentials
[\cite{Mewis19791,barnes1997tr,dullaert2005thixotropy}]. When
brought to rest, the properties of these systems may continue to
evolve as thermally driven particle motion drives structural
rearrangements
[\cite{Willenbacher_Aging1996,Cipelletti2000,Bonn_PRE2001,Kob_rearrangements2009,Osuji_Negi_PRE2009}],
a process known as physical aging. These rearrangements lead to
slowly increasing viscosity and complex modulus as the system
gradually ``relaxes'' and approaches some long time equilibrium or
steady state at rest. There is qualitative similarity between the
aging behavior of these colloidal systems and the glassy dynamics
observed in polymer and molecular structural glasses, key among them
the display of time-elapsed time rescaling of the dynamic properties
[\cite{Cloitre_PRL2000,cipelletti2002sdg,Cloitre_PRL2003,Joshi_PRE2008}].
However, non-trivial distinctions still separate the behavior of
colloidal systems from that of traditional glasses, as highlighted
recently [\cite{mckenna_JRheo2009}]. The viscosity of these fluids
under flow is comprised of both hydrodynamic and elastic or
thermodynamic contributions. In dilute dispersions of Brownian hard
spheres, based on Einstein's viscosity relation, the elastic and
viscous terms have been given by Batchelor
[\cite{Batchelor_brownian_stress_1977}] in Equations
\ref{eq:viscous_contributions},\ref{eq:elastic_contributions}

\begin{eqnarray}
\eta_r^v&=&1+2.5\varphi+5.2\varphi^2  \label{eq:viscous_contributions} \\
 \eta_r^e&=&0.97\varphi^2
\label{eq:elastic_contributions}
\end{eqnarray}

where $\eta_r^v$ is the viscous contribution to the relative
viscosity, $\eta_r^e$ is the elastic (Brownian) contribution and
$\varphi$ is the volume fraction of particles in the fluid. More
recent models have extended these considerations into concentrated
Brownian dispersions [\cite{Brady_conc_brownian1993}] as well as
weakly aggregated systems [\cite{Potanin_modeling1995}]. In
flocculating systems, the elastic term depends on the strength of
inter-particle attractions due to the presence of load-bearing
flocs, the size of which varies with the shear stress applied to the
system
 [\cite{COO:SonntagRussel1987_Experiment,COO:SonntagRussel1987_Theory,COO:PRE2008}].

Deconvolution of the viscosity into rate dependent elastic and
viscous terms can be made via stress jump measurements where a
suspension in steady state flow is subjected to a sudden drop to
zero shear rate, $\dot\gamma=0$. The viscous component of the shear
stress dissipates nearly instantaneously, leaving only the elastic
component as the measurable quantity that contributes to the
viscosity decay function, $\eta_-(t,\dot\gamma)$
 [\cite{Watanabe1996nonlinear}]. Suitable modification of strain
controlled rheometers employing a force rebalancing transducer for
torque measurements have enabled observations of the elastic stress
component at times as early as 20 ms after cessation of flow
 [\cite{Mackay_stress_jump_instrument_1992,COO:MacKayUnderwood1997_1,COO:Mewis2005_1}]
. Extrapolation of the time dependence of the stress back to the
cessation of shear at $t=0$ allows the steady state flow curve to be
represented in terms of the rate dependent elastic and viscous
contributions to the system viscosity as the shear stress before the
jump is given as
$\sigma(\dot\gamma)=\sigma^v+\sigma^e=(\eta^v+\eta^e)\dot\gamma$.
The difficulty of performing these measurements is matched by the
insight they provide into the dispersion rheology. For example, in
flocculating materials, the absence of hydrocluster induced shear
thickening has now been attributed to the precipitous decrease in
the elastic component of the shear stress with increasing shear
rate, compared to the rather more modest increase of the
hydrodynamic component [\cite{COO:Zukoski2004_1}].

In addition to decoupling the flow stress components, stress jump
measurements also permit the observation of the relaxation of the
elastic stress. A single exponential form was observed in weakly
flocculated colloidal dispersions of fumed silica and carbon black
[\cite{COO:Mewis2005_1}] out to $\approx 0.1$ s, whereas a power-law
dependence was found in hard sphere systems
[\cite{COO:MacKayUnderwood1997_1,Mackay_Langmuir2000}] out to
$\approx 1$ s. Stress jump measurements have also been applied in
polymer systems to follow the relaxation of deformed interfaces in
immiscible blends of poly(isobutylene) in poly(dimethylsiloxane)
(PDMS), where an exponential dependence was found, out to 4 s
[\cite{Mewis_stress_polymer1997}]. In more recent work on
polymer-clay nanocomposites, single exponential relaxations of the
elastic flow stress component were observed, both for a Newtonian
suspending polymeric fluid [\cite{Carreau_Mobuchon2007}] as well as
for a low molar mass PDMS medium [\cite{Mewis_PDMS_clay2009}], both
out to $\approx$ 0.2 s. Clearly, the dynamics of the elastic stress
displayed on cessation of shear, the residual stress, so to speak,
can provide important insights into the microscopics of the system
under study. In materials which display physical aging, the same
thermally driven structural rearrangements which give rise to
increasing moduli are also responsible for the decay of the residual
stress. Thus, one should expect that the time evolution of the
system properties should be related in some way to the dynamics of
this stress relaxation. Surprisingly, this connection remains quite
unexplored and residual stress dynamics in colloidal systems have
typically been considered only at short times scales, versus the
comparatively long times at which physical aging is studied.

Here, we perform stress jump measurements on a Laponite dispersion
and monitor the relaxation of the residual stress out to 100 s, as a
function of the rate of flow arrest. We also study the physical
aging of the system via small amplitude strain controlled dynamic
tests following cessation of shear. Our data reveal striking
parallels between the physical aging and residual stress relaxation
dynamics. In particular, the characteristic time for the stress
relaxation increases with the duration of the flow quench step.
Concurrently, the rate of physical aging also decreases, implying a
frustrated approach to the long time equilibrium state of the
system.

\section{\label{experimental}Experiment}
\subsection{\label{materials}Materials}
Our system is an aqueous dispersion of Laponite which consists of
disc-like clay particles of roughly 25 nm diameter and 1 nm
thickness. Repulsive glasses are formed at low ionic strength in
fairly dilute conditions due to long range electrostatic repulsion
between the particles~\cite{bonn_epl_1998}. The system displays
glassy dynamics, as studied by scattering
[\cite{knaebel2000abl,Bonn_PRE2001,ranjini_mochri_2004}] and
rheological means
[\cite{Bonn_Tanaka_rejuvenation_PRL2002,Bonn_JRheol2003,Joshi_PRE2008}].
Samples of concentrations $\varphi=$ 3, 4 and 5 wt.\% are prepared
by dispersion of Laponite XLG powder (Southern Clay Products) into
ultra-pure water, adjusted to pH 9.5 via addition of NaOH to ensure
chemical stability of the particles [\cite{mourchid_PRE_1998}]. The
samples are agitated vigorously for 20 minutes and then allowed to
develop completely for 5 days under quiescent conditions, defining a
well controlled and reproducible initial state.

\subsection{\label{methods}Method}

Rheology is conducted in strain controlled mode using an MCR301
instrument (Anton-Paar) in the cone-plate geometry (1$^{\circ}$, 50
mm diameter steel cone). Evaporation of water from the sample was
successfully suppressed without perturbing our system via
application of a thin film of mineral oil at the sample edge. Steady
state flow curves were collected from $\dot\gamma=10^{-2}$ to 10$^4$
s$^{-1}$ using a logarithmically varying equilibration time from 60
s at the lowest shear rates to 1 s at the highest shear rates.
Conventional stress relaxation data were taken after a resting
period of 300 s following a pre-shear at 100 s$^{-1}$, using a
strain of $\gamma=3\%$, which is within the linear viscoelastic
regime for all compositions studied. Frequency sweeps were carried
out using this same strain from 100 to 1 rad/s.

\subsubsection{\label{pre-shear_variation}Variation of
pre-shear rate}

Samples were subjected to varying pre-shear flows at rates between 4
and 1000 $s^{-1}$ for a duration of 100 s. In each case, this time
was more than sufficient to establish a robust steady-state in which
the viscosity of the sample was well defined within $\pm 5\%$. The
samples were subjected to a linear quench of 1 s duration ($t_q$ = 1
s) from the flowing state to the stationary or at-rest state.
Physical aging and stress relaxation measurements were carried out
on the system following separate pre-shear and mechanical quenching
steps as just described. The physical aging of the samples after
cessation of flow was monitored via the time evolution of the
complex shear modulus in the linear viscoelastic regime using strain
controlled oscillations at $\gamma=3$ \% at an angular frequency of
$\omega=10$ rad/s. The relaxation of the residual elastic component
of the flow stress in the system after cessation of flow, here
termed the residual stress, was then characterized, as described in
a prior publication [\cite{COO:RheolActa2009}]. Briefly, a constant
zero strain rate condition was imposed on the sample. The stress
required to maintain this $\dot\gamma=0$ stationary condition is the
residual stress. Checks were performed using a Newtonian fluid,
which should have no measurable residual stress, to verify the
ability of the instrument to accurately make the measurements
reported here.

\subsubsection{\label{pre-shear_variation}Variation of
quench duration} Samples were pre-sheared at a fixed rate of
$\dot\gamma=100$ s$^{-1}$ for 100 s and then brought to rest via a
linear ramp over varying durations, $t_q$, from 1 to 300 s.
Following this cessation of flow, physical aging was monitored as
above via the time evolution of the complex modulus in the linear
viscoelastic regime, $\gamma=3.0\%$ and $\omega=10$ rad/s. Following
the same pre-shear and flow cessation, the relaxation of residual
stresses were monitored, as described above.

\section{\label{results_discussion}Results and Discussion}

\subsection{\label{pre_shear_effects}Effects of Pre-Shear Rate}

\begin{figure}[h]
\includegraphics[width=70mm, scale=1]{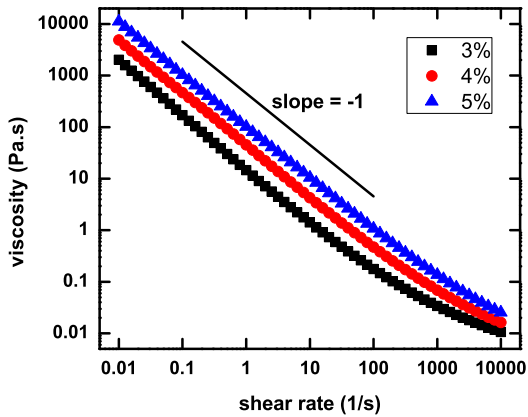}
\caption{Flow curves for 3, 4 and 5 wt.\% samples on decreasing
shear rate. \label{flow_curves}}
\end{figure}

Laponite forms a yield stress fluid when dispersed in water, even at
relatively low concentrations. As a result, the steady flow curves
do not display Newtonian plateaus at low shear rates, as shown in
Figure \ref{flow_curves}. The fluids display monotonic shear
thinning, with shear thinning exponents $|n|>0.9$ for all three
concentrations considered here. We do not observe high shear rate
viscosity plateaus, even out to $10^4$ $s^{-1}$. This strong shear
thinning behavior over a wide range of shear rates is typical of
Laponite suspensions
[\cite{Bonn_Tanaka_rejuvenation_PRL2002,Bonn_JRheol2003}].
Additionally, we do not observe any stress plateaus in the flow
curves, suggesting that there are no obvious intervening regimes of
shear banding or other flow-induced structuring. Absence of such
makes these suspensions ideal candidates for studies as conducted
here.

\begin{figure}[h]
\includegraphics[width=70mm,
scale=1]{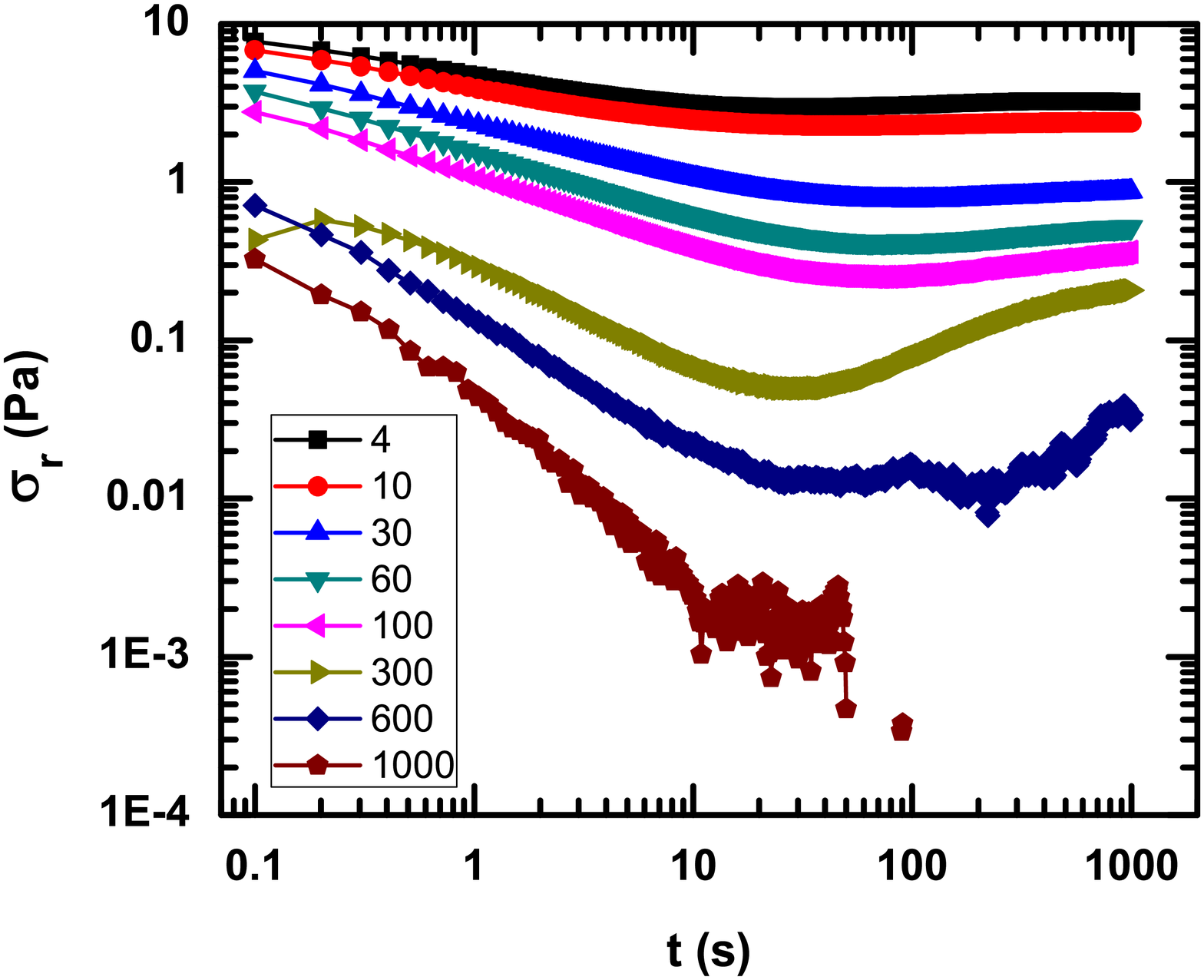}\\
\includegraphics[width=70mm,
scale=1]{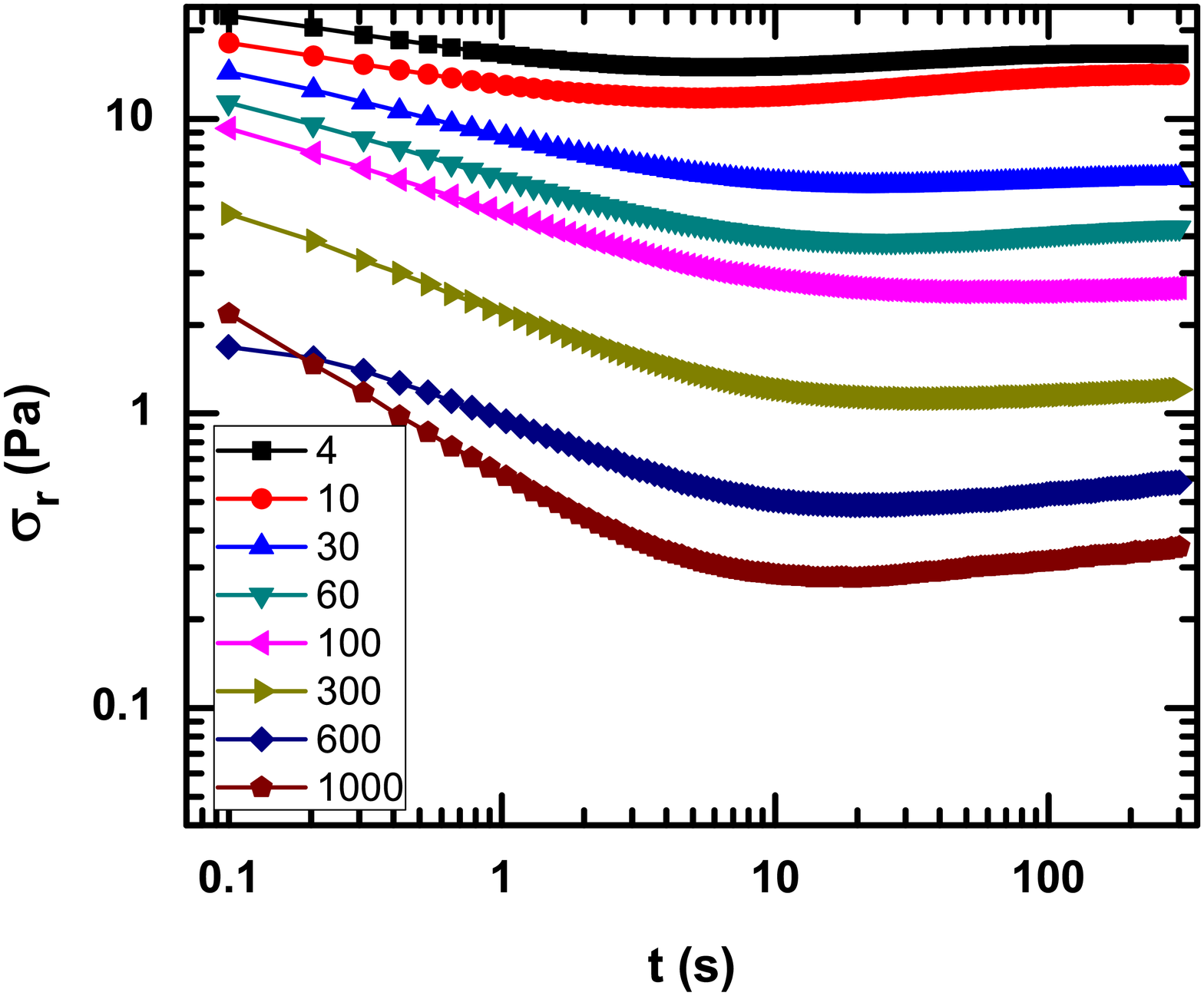}\\
\includegraphics[width=70mm, scale=1]{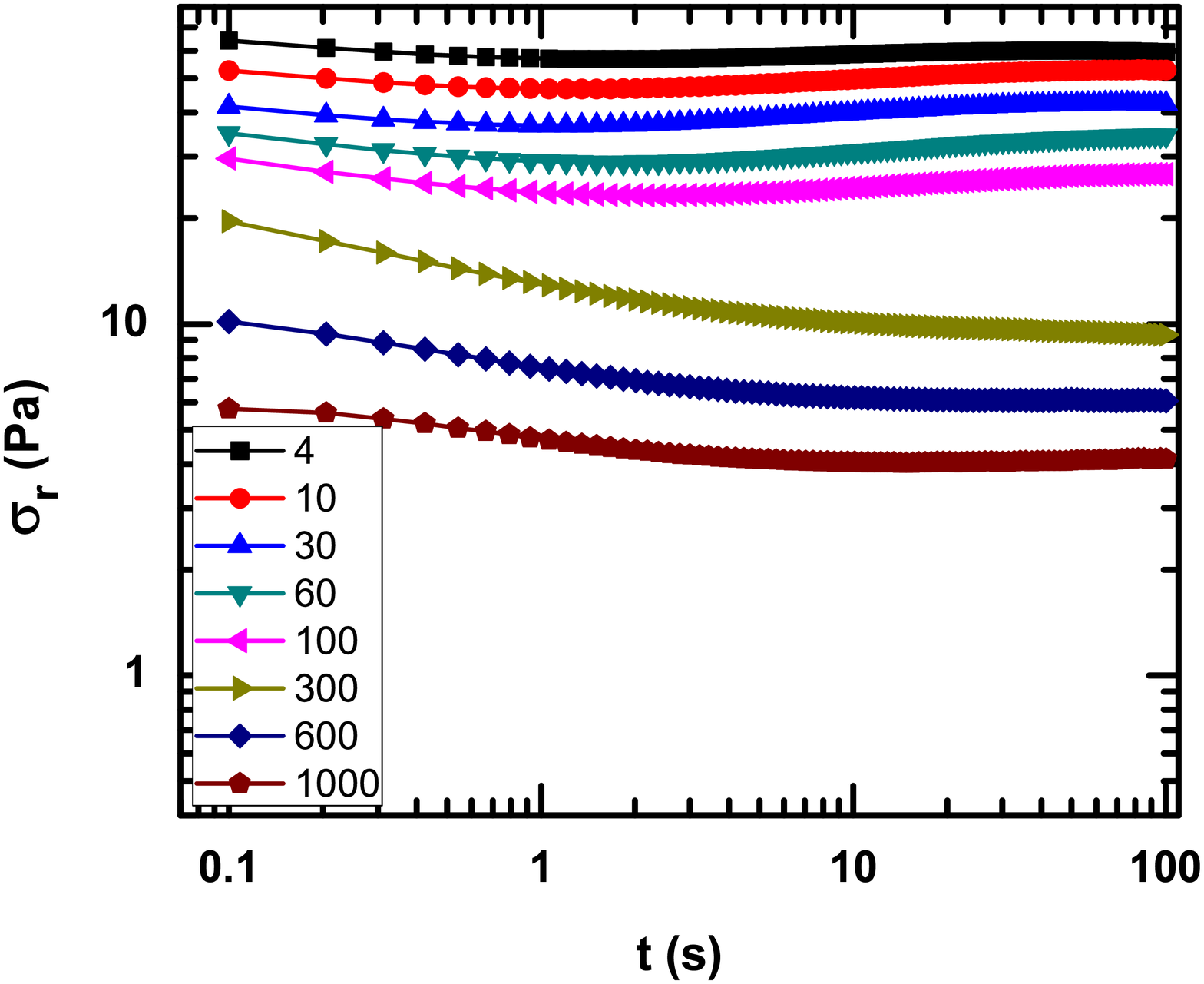}
\caption{Relaxation of residual stress or elastic stress component
for 3, 4 and 5 wt.\% samples (top to bottom) following stress jump from different
shear rates on the flow curve as indicated.
\label{internal_stress_relaxation1}}
\end{figure}

\begin{figure}[h]
\includegraphics[width=70mm, scale=1]{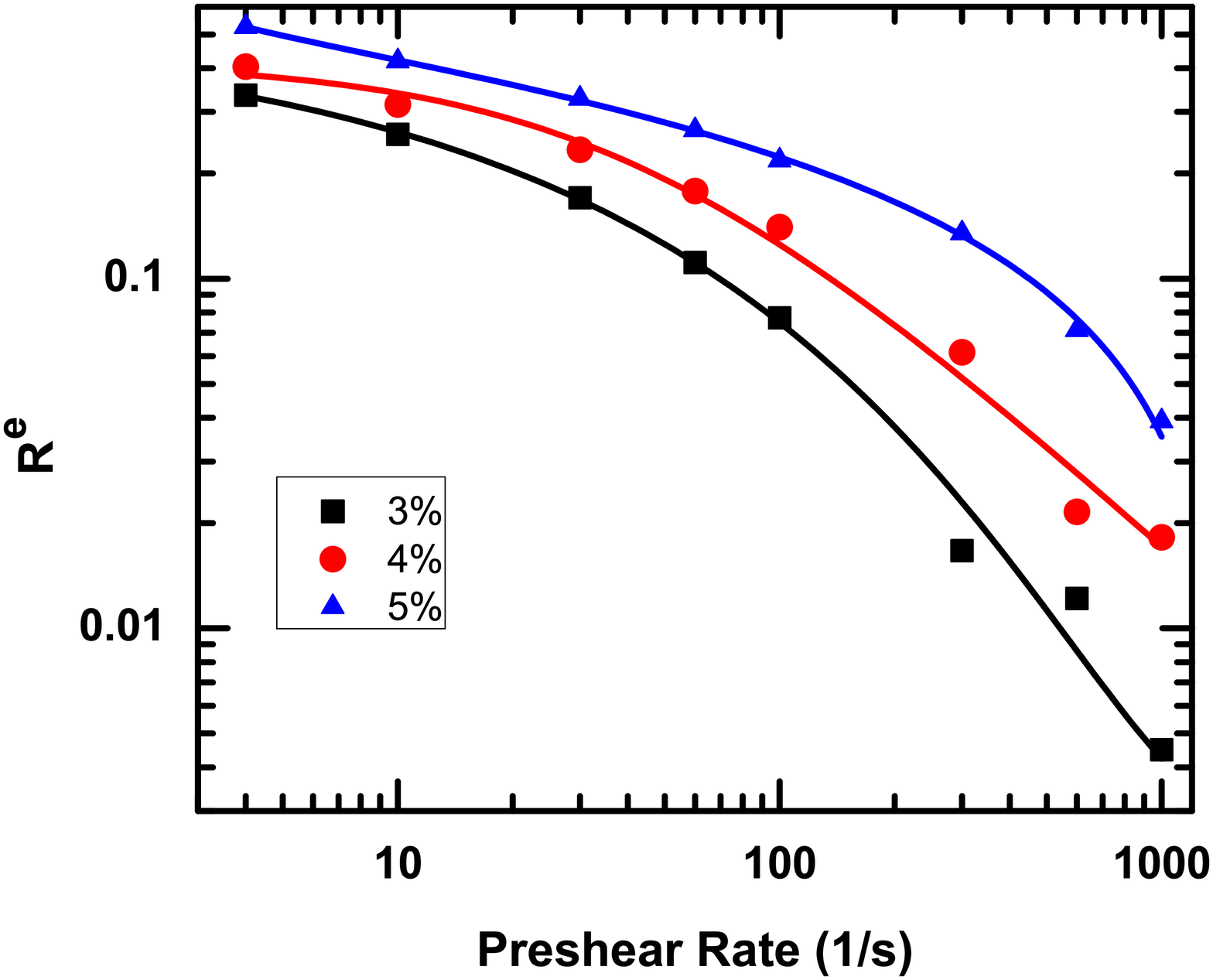}
\caption{Ratio of elastic stress at t=0.2 s to the overall flow
stress as a function of the shear rate. Lines are drawn as a guide
to the eye.\label{elastic_component}}
\end{figure}

Stress relaxation measurements on cessation of flow were performed
for shear rates between 4 and 1000 s$^{-1}$, as shown in Figure
\ref{internal_stress_relaxation1}. The data show a slowing rate of
decay of the residual stress as the initial shear rate of the system
is decreased. In some cases, at longer times, there are slight
increases in the stress which are attributed to structuring of the
system at rest, which causes the viscosity of the suspension to rise
with time. Because the rheometer in practice cannot maintain an
identically zero shear rate (in general the minimum controllable was
$\dot\gamma\approx$ 10$^{-6}$ s$^{-1}$), this results in an
increasing component of the stress that may become measurable at
long times. The data for 3 wt.\% following cessation from
$\dot\gamma=300$ s$^{-1}$ are particularly egregious however, for
reasons that remain unclear. The instantaneous value of the residual
stress immediately upon flow cessation is required to accurately
determine the elastic component of the flowing stress at different
points on the flow curve. Our measurements, however, did not permit
observation of the residual stress at times shorter than 0.1 s. We
take $\sigma_r(t=0.2$ s$)$ as representative of the elastic stress
component, $\sigma^e$, during the immediately preceding flow, rather
than performing a log-linear regression of the data to $t=0$, which,
strictly speaking is required for this purpose. Performing a
log-linear regression would imply the assumption that the same
functional form is maintained at shorter times, which is something
that we cannot justify with complete certainty. Despite the
quantitative difference between the two estimates, the data do
provide the same qualitative result. The elastic stress component
can be expressed as a fraction of the overall flow stress as
$R^e(\dot\gamma)=\sigma_r(\dot\gamma;t\approx
0)/\sigma(\dot\gamma)$. As shown in Figure \ref{elastic_component},
the elastic stress component decreases dramatically with increasing
shear rate, illustrating the strong reduction in the cohesion of
particle assemblies under flow. While this approach results in an
underestimate of the elastic stress component, it does capture the
large decrease in $R^e$ with increasing shear rate that is expected
in these systems.

\begin{figure}[h]
\includegraphics[width=70mm,
scale=1]{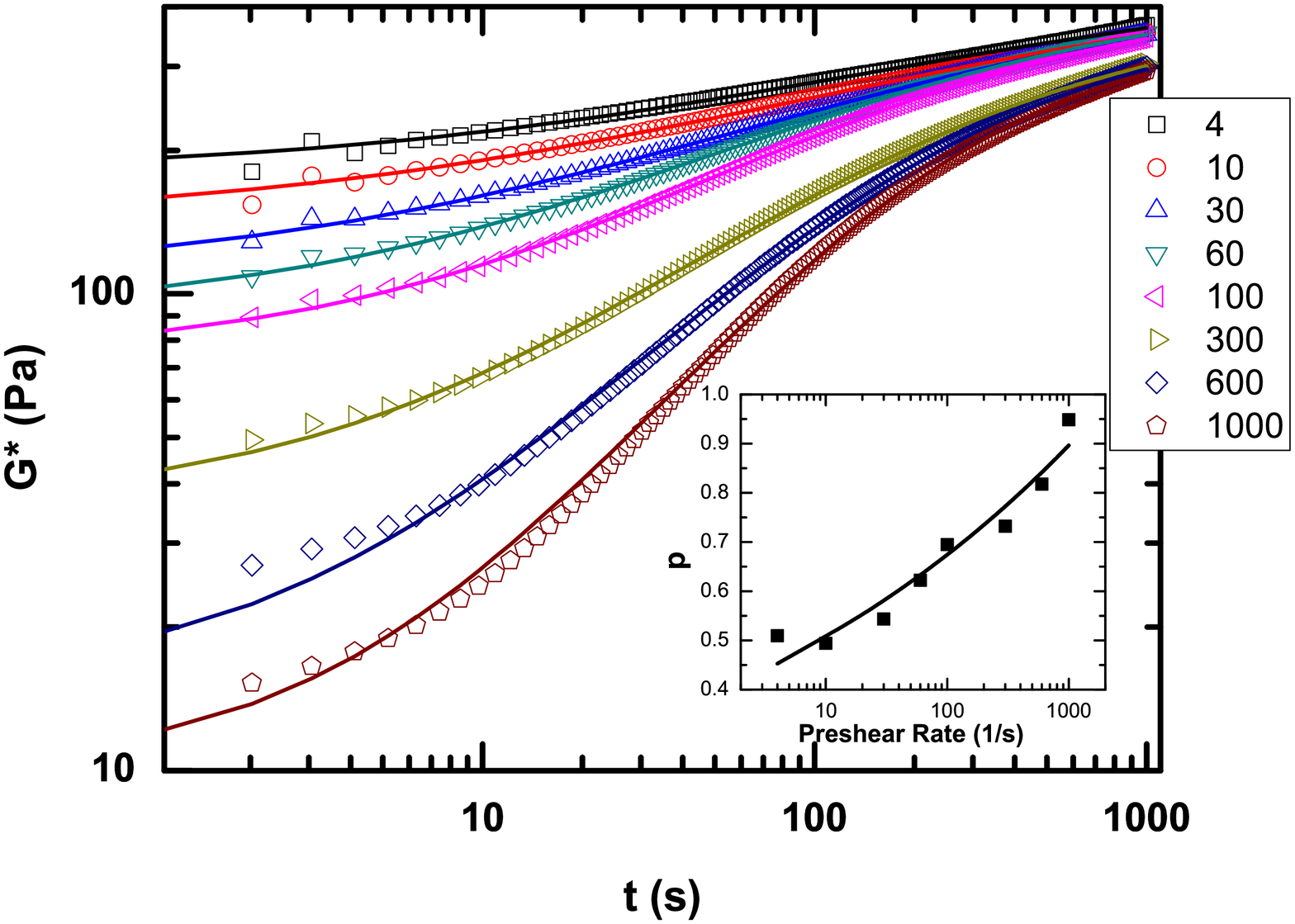} \\
\includegraphics[width=70mm,
scale=1]{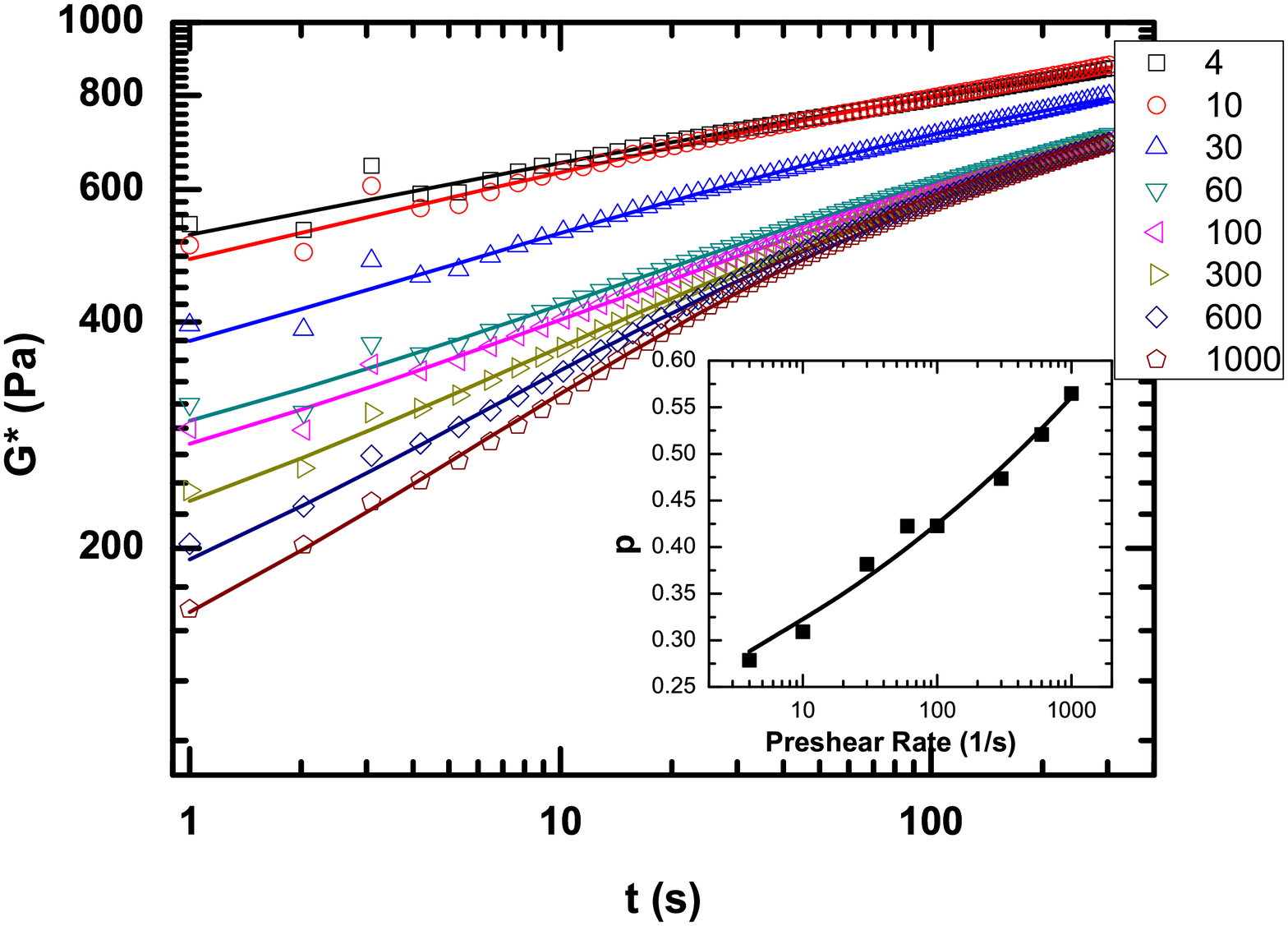} \\
\includegraphics[width=70mm, scale=1]{figures11242009/Fig4b_diff_preshear_complex_mod_4.eps}
\caption{Evolution of the complex modulus for 3, 4 and 5 wt.\%
samples (top to bottom) following stress jumps from different shear rates on the
flow curve as indicated. The lines are fits to the data using
Equation \ref{eq:modulus_evolution_fit}. Oscillations were performed
using a strain of $\gamma=3$ \% at an angular frequency of 10 rad/s.
Inset: Value of the exponent $p$ in Equation
\ref{eq:modulus_evolution_fit} describing the slope of $G^*(t)$ in
the vicinity of $t=\tau$. Lines are drawn as a guide to the eye.
\label{modulus_evolution1}}
\end{figure}

Physical aging measurements following rate quenches from the same
flow rates as above show that at long times, all samples followed
what appeared to be a weak power-law evolution of the complex
modulus with time, Figure \ref{modulus_evolution1}. However, the
behavior at short times varied strongly with the initial shear rate,
and the vertical separation at long times increased with increasing
volume fraction. Quenches from the lowest shear rate, $\dot\gamma=4$
s$^{-1}$ produced a higher complex modulus, with $G'>G''$, at short
times which evolved with a monotonic near power law display
throughout. By contrast, quenches from the highest shear rates
$\dot\gamma=1000$ s$^{-1}$ produced systems that were initially
fluid-like, with $G''>G'$ at short times, for $\varphi=3$ wt.\%. The
complex modulus evolved in a sigmoidal fashion, rapidly increasing
at intermediate times before assuming the weak near power law growth
at long times. Quenches from intermediate shear rates showed a
smooth variation between these two limiting behaviors. No
intersection of the different trajectories was observed during the
measurement.

The sigmoidal variation of the modulus in time is quite striking,
but not entirely unexpected. While many disordered colloidal systems
display power-law aging as implicit in time-elapsed time rescaling
[\cite{Cloitre_PRL2000,knaebel2000abl,COO:Cipelletti2005_slow_dynamics_1}],
intuitively, one understand that the power law dependence cannot
persist indefinitely, if it is assumed that the system actually has
a well defined, finite state or modulus at long time. Part of the
difficulty in studying these systems, however, is the slow nature of
the dynamics, and the slowing of the dynamics with the approach to
equilibrium or steady state. Here we see that by starting from a
highly fluid state characterized by a vanishing elastic stress
component, we are able to access the short time regime of the system
where it displays a sigmoidal response as it ages at first slowly,
then rapidly, then slowly again as it approaches its long time
steady state. This rapid evolution of the complex modulus is
mirrored in the rapid relaxation of the residual stress for samples
quenched from the highest shear rates. For a fixed shear rate,
moving to higher concentrations increases the residual stress
component and slows the dynamics of the system overall. A
Carreau-like model, Equation \ref{eq:modulus_evolution_fit} was used
to estimate a characteristic relaxation time for the sigmoidal
evolution of the complex modulus.

\begin{equation}
G^*(t)=G^*(\infty)+\frac{G^*(0)-G^*(\infty)}{\left[1+(t/\tau)^p\right]}
\label{eq:modulus_evolution_fit}
\end{equation}

The rate of aging can be represented by the parameter $p$ which
gives an indication of the steepness of the curve near the midpoint,
$t=\tau$, of the growth between presumed constant $t=0$ and
$t=\infty$ values. The inset plots of Figure
\ref{modulus_evolution1} show the increase of $p$ with increasing
shear rate before flow cessation, as described. Interestingly, a
sigmoidal response has also been observed in the time evolution of
the elastic modulus of a system composed of multi-arm star polymers
in solution [\cite{christopoulou2009ageing}] The soft colloidal
glass thus formed has important differences from the currently
considered Laponite suspension. Notably, aging in the polymer glass
proceeds via a complex two-step mechanism
[\cite{Helgeson2007,Rogers_Petekidis2010}], and the system is able
to access a presumed equilibrium state at long times ($\approx$ 5E4
s) where the modulus is stable in time. Nevertheless, the parallels
are noteworthy.

\begin{figure}[h]
\includegraphics[width=70mm, scale=1]{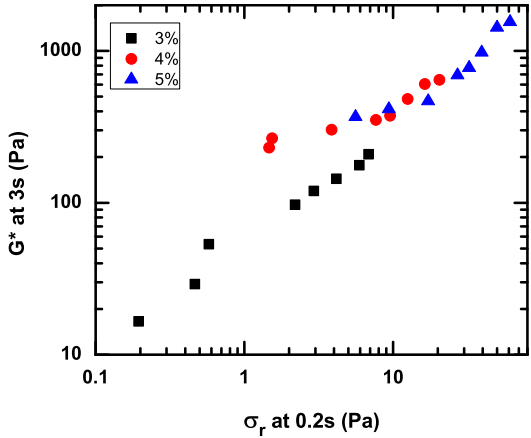}
\caption{Relation between the modulus and the stress on shear
cessation, for different pre-shear
rates.\label{modulus_stress_correlation}}
\end{figure}

The correlation between the earliest measured modulus of the system
and the short time value of the residual stress is shown in Figure
\ref{modulus_stress_correlation}. Smaller residual stresses are
produced by quenches from higher shear rates and are correlated to
lower values of the complex modulus. A linear dependence has been
observed between the elastic modulus and the residual stress in
particulate dispersions with attractive interactions
[\cite{COO:PRE2008,Ovarlez_PRE2008}]. Here, the dependence is
roughly linear, although for reasons that remain unclear, the low
stress data for $\varphi=4$ wt.\% are somewhat non-conforming.

\subsection{\label{pre_shear_effects}Effects of Quench Duration}

\begin{figure}[h]
\includegraphics[width=70mm, scale=1]{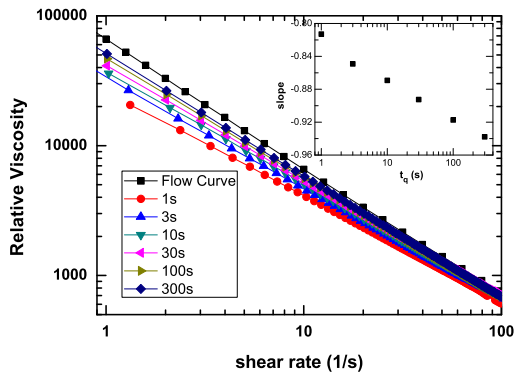}
\caption{Different quench rates result in the samples traversing
different paths in the ergodic to non-ergodic transition on
cessation of flow. Data are for a 4 wt.\% sample. Inset: Shear
thinning exponent during the quench decreases in magnitude as a
function of decreasing quench duration. \label{quench}}
\end{figure}

The preceding data clearly show that the dynamics of physical aging
are paralleled by those of residual stress relaxation. This is not
entirely surprising as the same thermally driven particle motion
that gives rise to rearrangements that age the system also result in
the relaxation of the elastic stresses built into the system on
cessation of flow. Additionally, aging from a more fluid state
should proceed faster than from a more jammed state. If we now
consider quenches from a fixed shear rate, but over varying
durations, we should expect to observe at some point the finite
effects of a non-steady state flow history on the sample. A rapid
flow cessation transitions the system from a free flowing steady
state to a stationary, jammed state in a short time. Increasing the
duration of the quench allows the system to sample configurations at
lower shear rates en-route to the stationary state. The equilibrium
condition at any particular shear rate is provided by the known
viscosity from the steady state flow curve and so the instantaneous
viscosity during the quench provides a good metric for the departure
of the system from equilibrium during the cessation of flow. Over a
range of quench durations, the system should display a spectrum of
behavior in which the viscosity at any given shear rate during the
quench deviates from the steady state value by an amount that scales
with the quench rate. We monitor the instantaneous viscosity during
the cessation of flow and find precisely this behavior, shown in
Figure \ref{quench}. As expected, the magnitude of the shear
thinning exponent, $n$ where $\eta(\dot\gamma)\sim\dot\gamma^n$
decreases continuously, in a linear fashion, as the quench duration
is decreased.

\begin{figure}[h]
\includegraphics[width=70mm,
scale=1]{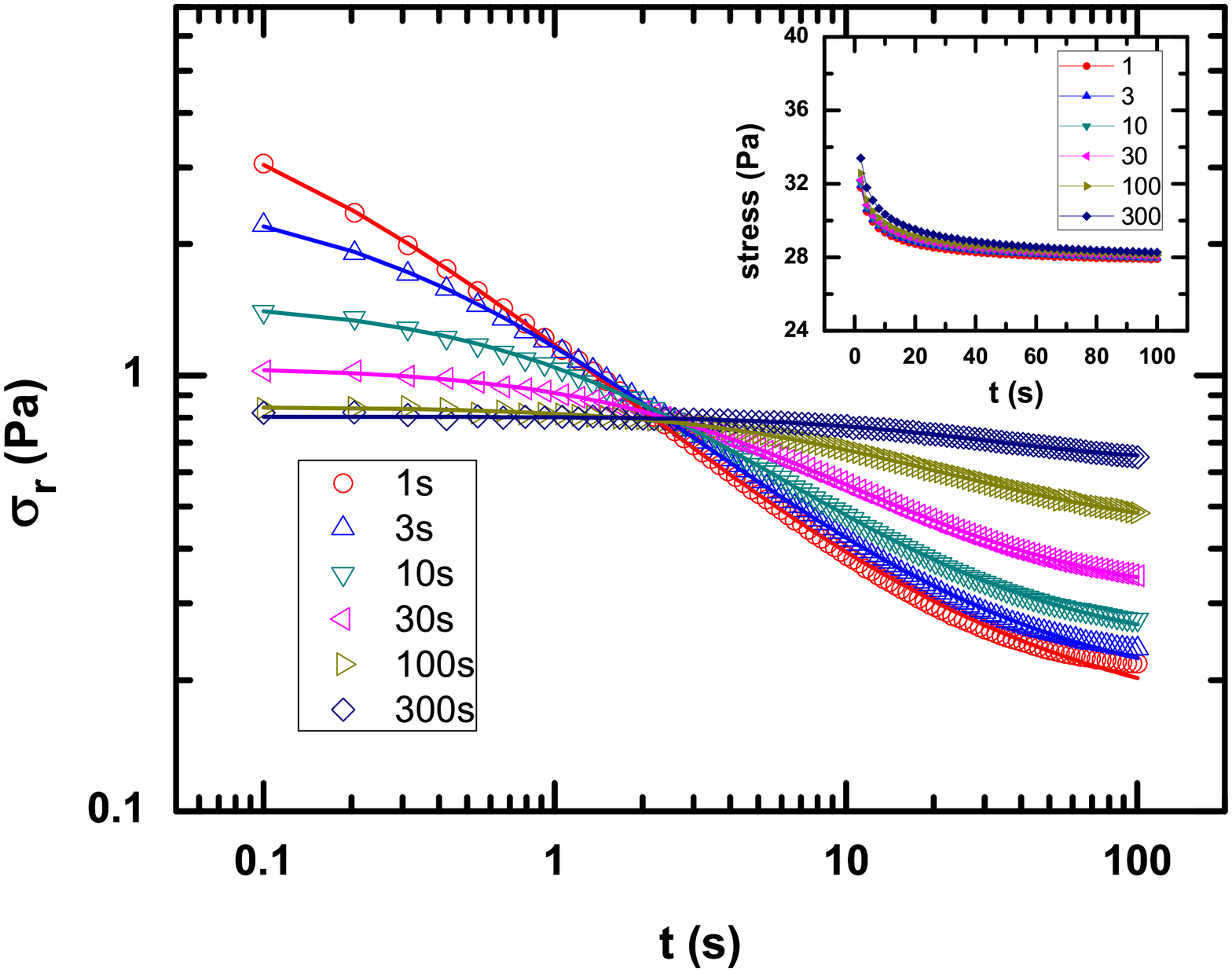} \\
\includegraphics[width=70mm,
scale=1]{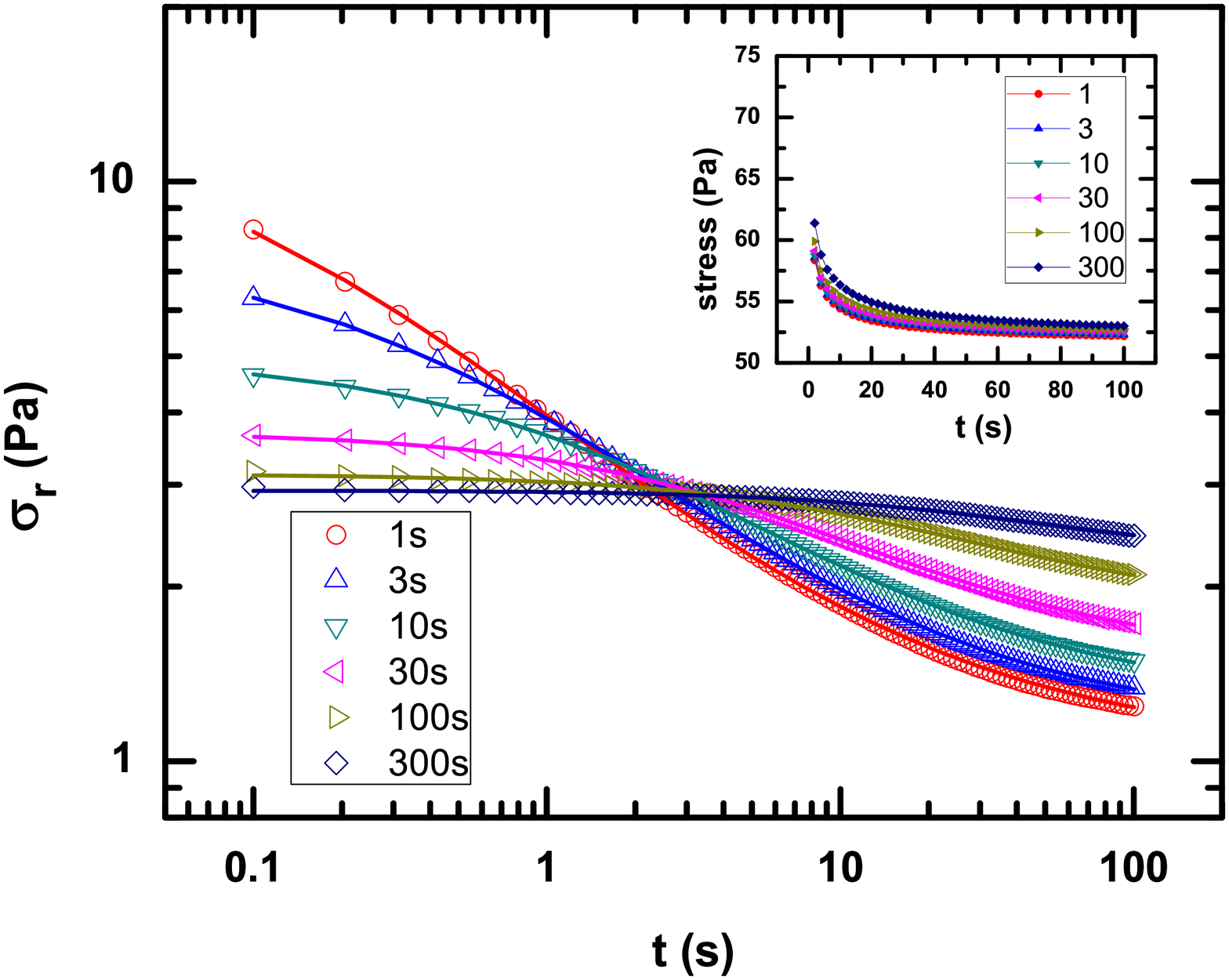} \\
\includegraphics[width=70mm, scale=1]{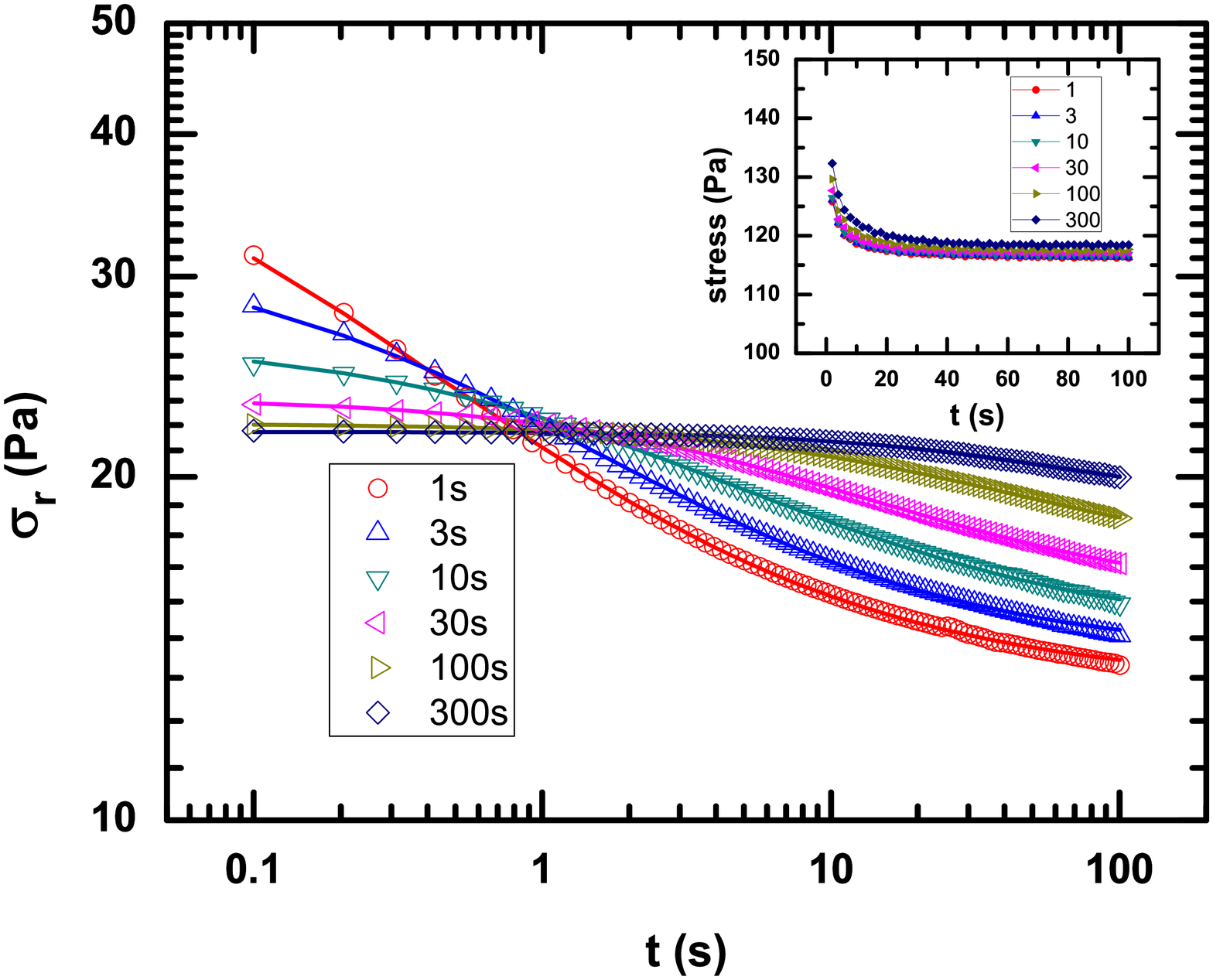}
\caption{Evolution of residual stress for different quench durations
as indicated, for 3, 4 and 5 wt.\% samples (top to bottom) quenched from 100
s$^{-1}$. Lines are fits to the data using Equation
\ref{eq:stress_relaxation_fit}. The insets show the flow stress
during the rejuvenation steps, indicating that the system starts
from a consistent state. \label{stress_relaxation}}
\end{figure}

Prior to measurement of residual stresses, as well as for the
modulus measurements, samples were subjected to a rejuvenating flow
at the selected shear rate, $\dot\gamma=100$ s$^{-1}$. During these
rejuvenating steps, the flow stress and thus the viscosity evolve to
well defined levels that are characteristic of the steady state of
the system at the applied flow rate (insets, Figure
\ref{stress_relaxation}).  The elastic stress that results on flow
cessation is due to the structuring of the system that takes place
during the finite quench interval. Surprisingly, we find that fast
quenches lead to a higher residual stress than produced for slow
quenches. Additionally, the residual stresses relax faster and to
lower values for fast quenches than those produced by slower
quenches, Figure \ref{stress_relaxation}. The data are strikingly
different in form than those for a conventional stress relaxation,
in response to a step-strain. They exhibit neither a power law nor
stretched exponential dependence on time, but instead are suggestive
of a sigmoidal approach to a finite, non-zero value at long times,
much like the residual stress data of Figure
\ref{internal_stress_relaxation1}. Notably, the relaxation shows a
clear and monotonic dependence on the quench duration, $t_q$.

\begin{figure}[h]
\includegraphics[width=70mm, scale=1]{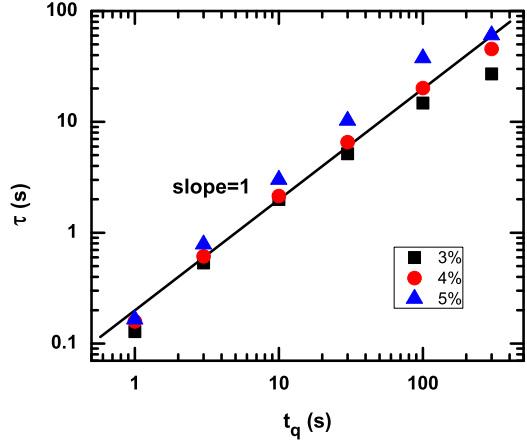}
\caption{Quench time dependence of the characteristic stress
relaxation time, $\tau$ derived from Equation
\ref{eq:stress_relaxation_fit}. The line shows a slope of 1.
\label{relaxation_times}}
\end{figure}

We fit our data with good fidelity again using a sigmoidal function,
Equation \ref{eq:stress_relaxation_fit} for the time dependence of
the residual stress, $\sigma_r(t)$. Here, again, $\tau$ is a
characteristic time that describes the decrease of the sigmoidal
function to its vertical midpoint between the value at $t=0$,
$\sigma(0)$ and at $t=\infty$, $\sigma(\infty)$, and the exponent
$p$ characterizes the slope of the function in the vicinity of
$t=\tau$. Remarkably, we find that the characteristic time for the
relaxation of the residual stress is a linear function of the quench
time, $\tau\sim t_q$, as shown in Figure \ref{relaxation_times}.

\begin{equation}
\sigma_r(t)=\sigma_r(\infty)+\frac{\sigma_r(0)-\sigma_r(\infty)}{\left[1+(t/\tau)^p\right]}
\label{eq:stress_relaxation_fit}
\end{equation}

\begin{figure}[h]
\includegraphics[width=70mm, scale=1]{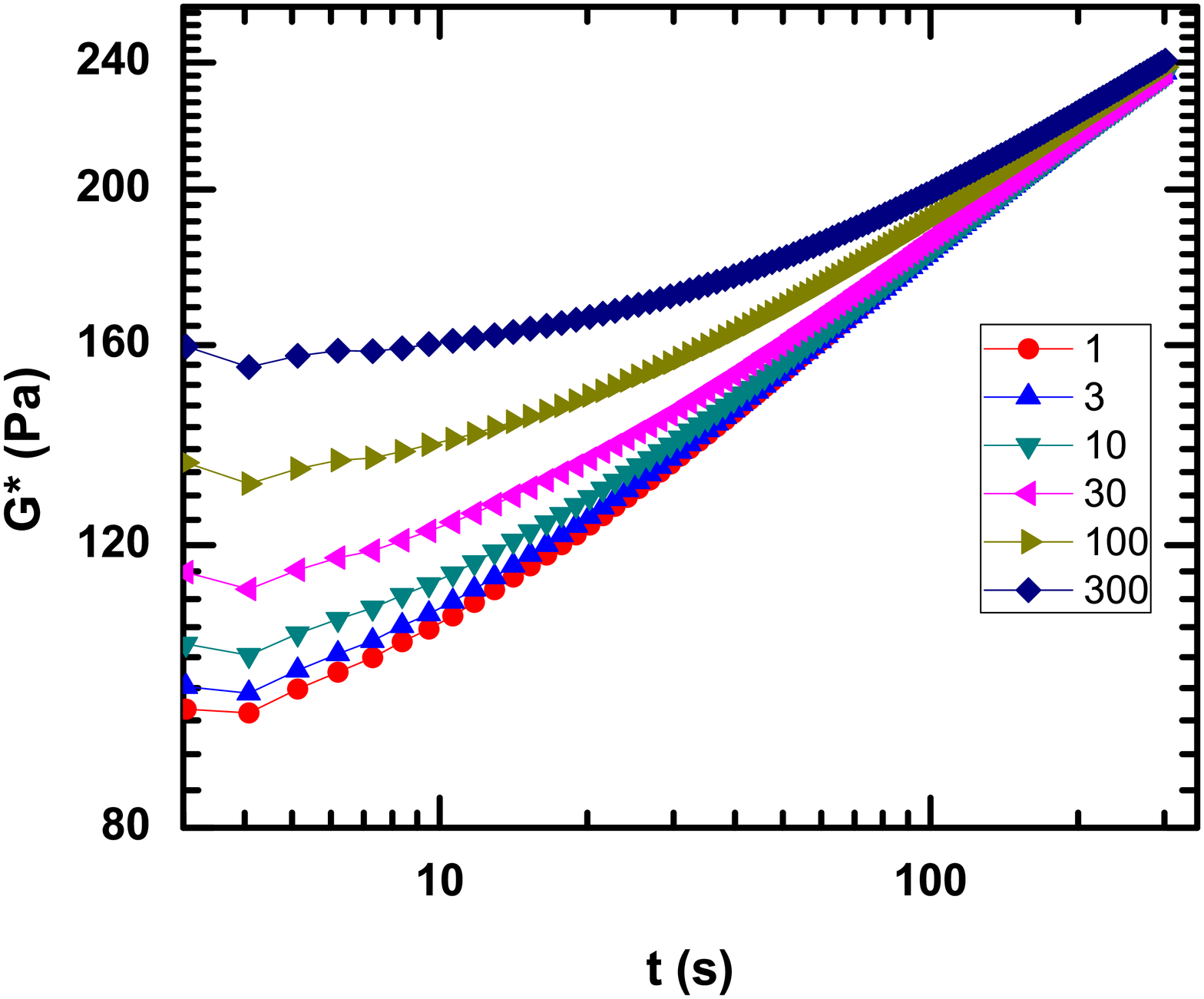} \\
\includegraphics[width=70mm, scale=1]{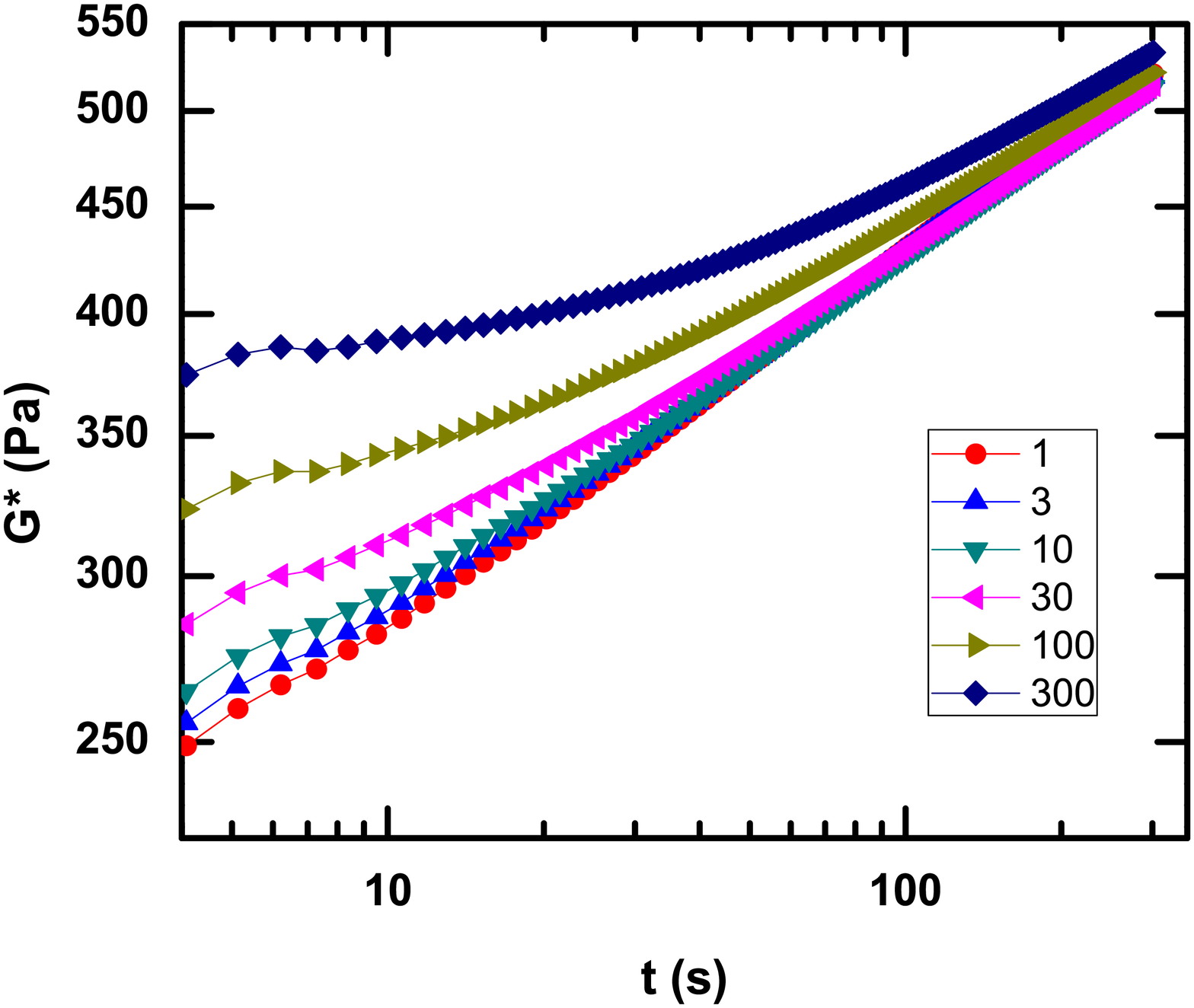} \\
\includegraphics[width=70mm, scale=1]{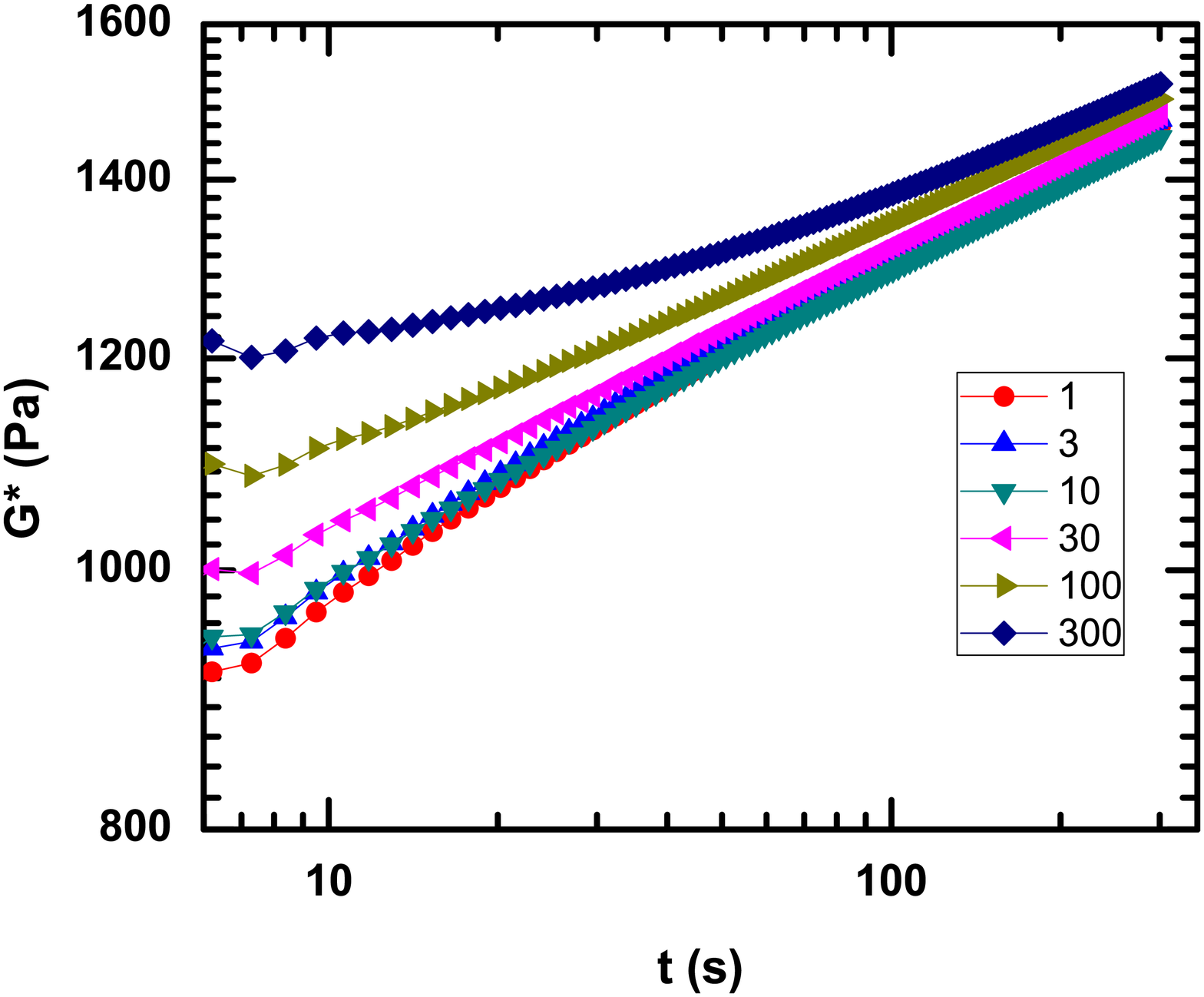}
\caption{Evolution of the complex modulus in 3, 4 and 5 wt.\%
samples (top to bottom) quenched from $\dot\gamma=100$ s$^{-1}$. Oscillations are
performed using a strain of $\gamma=3$ \% at an angular frequency of
10 rad/s. The legend indicates the duration of the quench, $t_q$, in
seconds. \label{modulus_evolution}}
\end{figure}

\begin{figure}[h]
\includegraphics[width=70mm, scale=1]{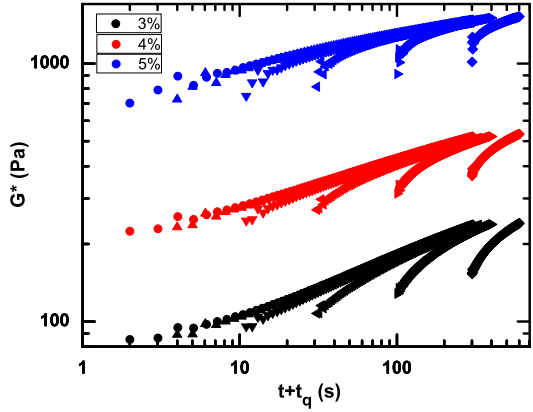}
\caption{Evolution of the complex modulus, shifted according to the
quench durations. \label{modulus_evolution_shifted}}
\end{figure}

\begin{figure}[htb]
\includegraphics[width=70mm, scale=1]{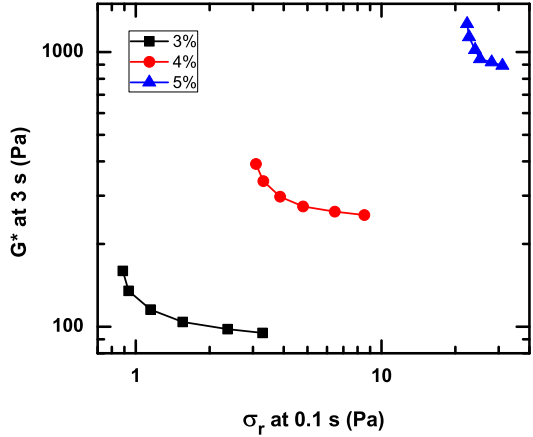}
\caption{Relation between the modulus and the stress on shear
cessation, for different quench
rates.\label{modulus_stress_correlation2}}
\end{figure}

The modulus of the glass, measured immediately after cessation of
shear is inversely proportional to the rate of quench. The slowest
quench rates, with flow cessation over periods of 300 s, lead to the
highest moduli. The modulus shows a slow increase that is well
described by a power law at late times, but with significant
sigmoidal character at shorter times,
Figure~\ref{modulus_evolution}. This is not unlike the response
observed for quenches from different initial shear rates. Attempts
to superpose the data by time-elapsed time rescaling fail, Figure
\ref{modulus_evolution_shifted}. That is, shifting in time to
account for the true sample age due to passage of time during the
quench does not produce a satisfactory superposition of data. These
results show that the flow history of the sample is important in
determining its aging path, and different trajectories through the
non-steady state flow curves en route to the stationary state
invariably result in different and non-superposable aging behaviors
thereafter. The correlation between modulus and residual stress is
shown in Figure \ref{modulus_stress_correlation2}. Here now, there
is no longer a simple direct correspondence between stresses and
moduli. The residual stress displayed is not just function of
structuring that occurs during the finite quench, but also
relaxation of elastic components that were supported at the start of
the quench.

\begin{figure}[h]
\includegraphics[width=70mm, scale=1]{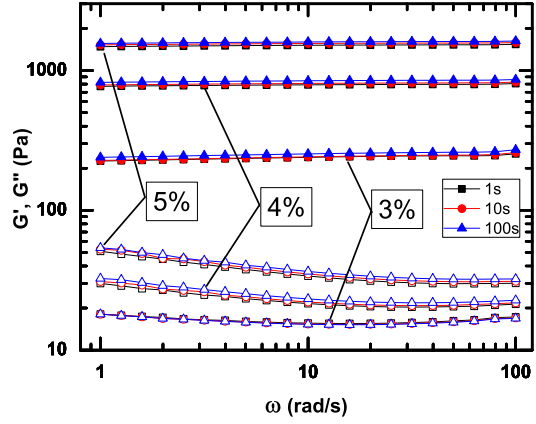}
\caption{Frequency sweep measurements. Samples were quenched over
the 3 different times indicated and aged for 300 s before
measurement. \label{frequency_sweep}}
\end{figure}

\begin{figure}[ht]
\includegraphics[width=70mm,
scale=1]{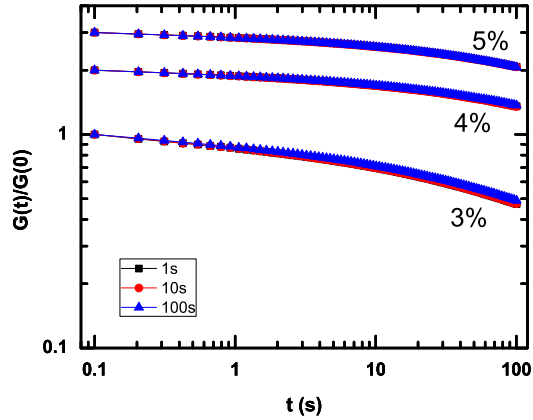}
\caption{Conventional stress relaxation measurements. Samples were
quenched over the 3 different times indicated and aged for 300 s
before measurement. \label{conventional_relaxation}}
\end{figure}

The data for the residual stress relaxation can be contrasted with
that obtained by traditional step-strain experiments as well as
frequency dependent moduli. Frequency sweeps of the suspensions show
a very weak dependence of both storage and loss moduli on frequency,
Figure \ref{frequency_sweep}. Data was collected after quenching
from $\dot\gamma=100$ s$^{-1}$ over 1, 10 and 100 s, following which
the samples were allowed to sit quiescently for 300 s prior to
measurement. This ensured that the measurement time ($\approx$ 100
s) was small enough compared to the sample age that changes in the
system properties due to aging during the measurement were
reasonably small. Similarly, step strain experiments were performed
after an 300 s quiescent period to evaluate the conventional stress
relaxation behavior. A strain of $\gamma=3\%$ was used, which is in
the linear regime for the system, as confirmed by strain sweep
experiments. The data, Figure \ref{conventional_relaxation}, are in
good agreement with the slow relaxation that is inferred from the
$G(\omega)$ data of Figure \ref{frequency_sweep}. Less than half a
decade of relaxation is observed over 3 decades of time for 3 wt.\%,
and substantially less than this for 4 and 5 wt.\%. For both
$G(\omega)$ and $G(t)$, there is little dependence on the quench
rate both in the absolute values of the moduli measured, and, more
significantly, in their time(frequency) dependences.

\section{\label{conclusion}Conclusion}

\begin{figure}[ht]
\includegraphics[width=70mm, scale=1]{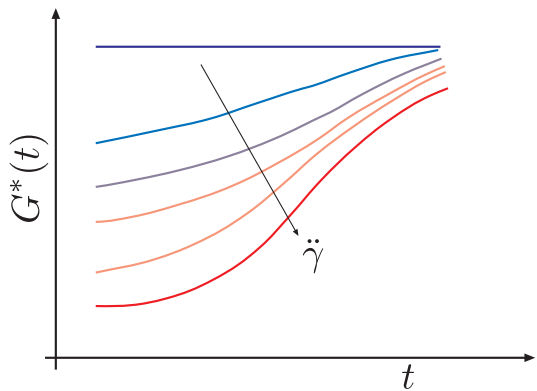}
\caption{Generic picture of the evolution of the complex
modulus.\label{generic_modulus_evolution}}
\end{figure}

In summary, our investigation has shown that the quench rate, as
well as the initial flow rate from which the system is quenched,
both strongly influence the aging dynamics. The effects of the
initial flow rate are particularly intuitive, assuming an infinitely
fast quench such that the properties (structure) immediately on
cessation of flow are those of the immediately preceding flow
regime. One can readily imagine that a system that is in steady flow
at an infinitely low shear rate, $\dot\gamma\rightarrow 0$, would
display an infinitesimal rate of aging upon cessation of flow.
Conversely, a system in steady flow at a very high shear rate,
$\dot\gamma\rightarrow\infty$, is completely fluidized, and should
display the the full aging response from fluid to semi-solid as the
system undergoes structural arrest and further vitrifies with
passage of time after cessation of flow. For all finite initial
shear rates, the displayed behavior should fall on a spectrum
between these two limiting cases, as described in Figure
\ref{generic_modulus_evolution}. Whether or not a power-law or
logarithmic or even exponential aging response is observed then
becomes a function of the initial flow rate of the system and the
duration of the observation window, among other factors.
Promisingly, recent work in this area has been successful in
revealing both the stress [\cite{Osuji_Negi_arrest_arXiv2009}] and
frequency [\cite{Osuji_Negi_arrest_arXiv2010}] dependence of the
complex modulus during structural arrest following rapid cessation
of flow from high shear rates. The behavior observed for the
currently studied Laponite system is in broad agreement with this
framework of a globally sigmoidal response which we advance as a
generic response for out of equilibrium materials. This argument has
also been given based on earlier work in polymer glasses, where a
sigmoidal form is suggested for a perfect thermal quench, with
power-law aging simply the consequence of a limited observation or
aging time window [\cite{mckenna2003mrp}]. Here, abrupt mechanical
quenching of the flowing suspension permits access to a larger
spectrum of the system response. The lack of superposition of the
complex modulus with waiting time may reflect the different energy
landscapes or distribution of relaxation mechanisms available to the
system as a function of the initial flow rate and the fact that the
evolution of the landscapes is not along a single trajectory, but
depends on the initial condition from which the system is evolved.

With a fixed initial condition, the quench rate is seen to exert a
strong influence on the aging dynamics in this repulsive colloidal
glass and serves to characterize the departure of the system from
equilibrium. Rapid quenches lead to faster aging from smaller
moduli. Further, the aging dynamics are well correlated to the decay
of residual stresses that are trapped in the glass at the ergodic to
non-ergodic transition on cessation of flow. These stress dynamics
are dependent on the quench rate and display a linear dependence of
their characteristic relaxation time on the duration of the quench.
Notably, they follow neither a simple exponential nor a power law
form. The simultaneously decreasing initial value and final extent
of relaxation of the residual stress concisely represent the
frustrated and continuously slowing approach to equilibrium in these
materials. This work suggests that the flow cessation rate may be
identified as somewhat akin to an effective cooling rate for the
colloidal glass wherein infinitely slow cooling,
$t_q\rightarrow\infty$, maintains equilibrium and results in a
stationary state on complete flow cessation. Many of the features
displayed here resonate intuitively, and should be observed in other
colloidal systems which display glassy or glass-like dynamics.
Dynamic scattering can provide useful data in these systems and
measurements which incorporate control over the flow arrest are a
natural follow on to this work.

\begin{acknowledgements}
The authors gratefully acknowledge NSF funding via CBET-0828905.
\end{acknowledgements}

\bibliography{internal_stress_aging}

\end{document}